\documentclass[openbib,12pt]{article}

\newcommand{\utwi}[1]{\mbox{\boldmath $ #1$}}

\usepackage{booktabs,dcolumn}
\usepackage{psfrag,graphicx}
\usepackage{eso-pic,graphicx}
\usepackage{amsmath}
\usepackage{graphicx}
\usepackage{pdflscape}
\usepackage{longtable}
\usepackage{epstopdf}
\usepackage{appendix}

\newcolumntype{z}[1]{D{.}{.}{#1}}

\newcommand{\cred}{\textcolor{red}}
\newcommand{\cb}{\textcolor{blue}}
\date{}

\begin{document}

\title{
\begin{center} {\Large \bf Semi-parametric Dynamic Asymmetric Laplace Models for Tail Risk Forecasting, Incorporating Realized Measures} \end{center}}
\author{Richard Gerlach, Chao Wang\\
Discipline of Business Analytics, Business School, \\The University of Sydney, Australia.}

\date{} \maketitle

\begin{abstract}
\noindent
The joint Value at Risk (VaR) and expected shortfall (ES) quantile regression model of Taylor (2017) is extended via incorporating a
realized measure, to drive the tail risk dynamics, as a potentially more efficient driver than
daily returns. Both a maximum likelihood and an adaptive Bayesian Markov Chain Monte Carlo method are employed for estimation, whose
properties are assessed and compared via a simulation study; results favour the Bayesian approach, which is subsequently employed in a
forecasting study of seven market indices and two individual assets. The proposed models are compared to a range of parametric, non-parametric
and semi-parametric models, including GARCH, Realized-GARCH and the joint VaR and ES quantile regression models in Taylor (2017). The
comparison is in terms of accuracy of one-day-ahead Value-at-Risk and Expected Shortfall forecasts, over a long forecast sample period
that includes the global financial crisis in 2007-2008. The results favor the proposed models incorporating a realized measure, especially
when employing the sub-sampled Realized Variance and the sub-sampled Realized Range.
\vspace{0.5cm}

\end{abstract}

\newpage
\pagenumbering{arabic}

{\centering
\section{\normalsize INTRODUCTION}\label{introduction_sec}
\par
}
\noindent

Since the introduction of Value-at-Risk (VaR), via J.P. Morgan's RiskMetrics model in 1993, more and more financial institutions worldwide
employ VaR to assist their capital allocation and risk management practice. VaR is a quantile of the conditional distribution of
financial returns and is a standard measurement for regulatory capital allocation. However, VaR attracts criticism since it says nothing
about losses beyond the quantile, i.e. for violating returns, and is also not mathematically coherent, in that it can
favour non-diversification. Expected Shortfall (ES), proposed by Artzner \emph{et al.} (1997, 1999), gives the expected
loss, conditional on returns exceeding a VaR threshold, and is a coherent measure, thus in recent years it is more widely employed for
tail risk measurement and now also recommended by the Basel Committee on Banking Supervision.

Although ES is widely used by financial institutions, its estimation remains a significant challenge, partly influenced by
ES not being elicitable (Gneiting, 2011), i.e. it lacks of a loss function that is uniquely optimised by its true value. However, recently
it is found that VaR and ES are jointly elicitable, by Fissler and Zeigel (2016), i.e. there is a joint loss function that is minimised by
the true values of VaR and ES considered jointly. Taylor (2017) exploited this class of joint loss functions to build some semi-parametric
time series models for VaR and ES; those models are extended in this paper, to incorporate realized measures.

There are many approaches to forecasting both VaR and ES: including parametric approaches, often based on the GARCH or Stochastic
Volatility (SV) models; semi-parametric approaches, including those based on Extreme Value
methods (e.g. Embrechts) or those based on quantile regression type models, e.g. the CAViaR of Engle and Manganelli (2004); and
non-parametric approaches, a popular example being historical simulation. The primary goal of this paper is the extend the
recent semi-parametric approach by Taylor (2017), who employs the Asymmetric Laplace density (ALD) to build a likelihood function whose
maximum likelihood estimates (MLEs) coincide with those obtained by the minimisation of a joint loss function for VaR and ES.

Realized measures of volatility play a key role in calculating accurate VaR or ES forecasts, and several realized measures
employing high frequency intra-day data are developed in the literature, going back to Parkinson (1980) and Garman and Klass (1980),
who proposed the daily high-low range as a more efficient volatility estimator, compared to the daily squared return; while more
recently several popular and efficient realized measures, including Realized Variance (RV): Andersen and Bollerslev (1998),
Andersen \emph{et al.} (2003) and Realized Range (RR): Martens and van Dijk (2007), Christensen and Podolskij (2007) are popular.
The availability of these measures has also inspired developments in parametric modelling of volatility and
semi-parametric modelling of quantiles and expectiles, where such measures are used as inputs, e.g. see Gerlach and Wang (2016), and
also employed to capture the contemporaneous relationships between the measure and the latent volatility, e.g. as in
Hansen \emph{et al.} (2011)'s Realized GARCH model.

In this paper, the model of Taylor (2017), denoted ES-CAViaR, is expanded to allow a realized measure to drive the tail risk dynamics.
The proposed model framework is called ES-CAViaR-X. Further, an adaptive Bayesian MCMC algorithm is utilised for estimation and forecasting
in the proposed model. To evaluate the performance of the proposed ES-CAViaR-X model, employing various realized measures as inputs,
the accuracy of the associated VaR and ES forecasts are assessed via an empirical study. Over the forecast period 2008-2016, results
illustrate that ES-CAViaR-X models perform favourably, compared to Taylor's ES-CAViaR model, and to a range of traditional competing models.

The paper is organized as follows:
A review of the ES-CAViaR type models is conducted in Section \ref{new_es_caviar_model}. Section \ref{model_section} formalizes
the ES-CAViaR-X class of models; The
associated likelihood and the adaptive Bayesian MCMC algorithm for parameter estimation are presented in
Section \ref{beyesian_estimation_section}. The simulation and empirical studies are discussed in Section \ref{simulation_section} and
Section \ref{data_empirical_section} respectively. Section \ref{conclusion_section} concludes the paper and discusses future work.

{\centering
\section{\normalsize ES-CAViaR MODELS}\label{new_es_caviar_model}
\par
}

%
%

Koenker and Machado (1999) noted that the usual quantile regression estimator is equivalent to a maximimum likelihood estimator when assuming that the data are conditionally Asymmetric Laplace (AL)
with a mode at the quantile, i.e. if $r_t$ is the data on day $t$ and $Pr(r_t < Q_t | \Omega_{t-1}) = \alpha$ then the parameters in the model for $Q_t$ can be estimated using a likelihood based on:
$$ p(r_t| \Omega_{t-1}) = \frac{\alpha (1-\alpha)}{\sigma} \exp \left( -(r_t-Q_t)(\alpha - I(r_t < Q_t)  \right)\,\, , $$
for $t=1,\ldots,n$ and where $\sigma$ is a nuisance parameter.

Taylor (2017) extended this result to incorporate the associated ES quantity into the likelihood expression, noting a link between $ES_t$ and a dynamic $\sigma_t$, resulting in the conditional density function:
$$ p(r_t| \Omega_{t-1}) = \frac{\alpha (1-\alpha)}{ES_t} \exp \left( -\frac{(r_t-Q_t)(\alpha - I(r_t < Q_t)}{\alpha ES_t}  \right)\,\, , $$
allowing a likelihood function to be built and maximised, given model expressions for $Q_t, ES_t$. Taylor noted that the negative logarithm of the resulting likelihood function is strictly consistent
for $Q_t, ES_t$ considered jointly, i.e. it fits into the class of strictly consistent functions for VaR, ES jointly developed by Fissler and Zeigel (2016).

Taylor (2017) considered two well-known model forms for $Q_t$, from Engle and Manganelli (2004)'s CAViaR framework, then further proposed two novel model forms for ES, both of which also describe the
dynamics between VaR and ES and specifically avoid ES estimates crossing the corresponding VaR estimates, as presented in Models (\ref{es_caviar_ar_model}) (ES-CAViaR-AR: ES-CAViaR with an autoregressive ES component)
and (\ref{es_caviar_exp_model}) (ES-CAViaR-Exp: ES-CAViaR with an exponential ES component):

\begin{eqnarray} \label{es_caviar_ar_model}
Q_{t}&=& \beta_0+ \beta_1 |r_{t-1}| + \beta_2 Q_{t-1},\\ \nonumber
 \text{ES}_t&=&Q_t-x_t, \\ \nonumber
 x_t&=&
\begin{cases}
    \gamma_0 + \gamma_1 (Q_{t-1} - r_{t-1}) + \gamma_2 x_{t-1} & \text{if } r_{t-1} \leq Q_{t-1},\\
    x_{t-1}              & \text{otherwise},
\end{cases}
\end{eqnarray}
where $\gamma_0 \ge 0, \gamma_1 \ge 0, \gamma_2 \ge 0$ ensure that the VaR and ES series do not cross.

\begin{eqnarray} \label{es_caviar_exp_model}
Q_{t}&=& \beta_0+ \beta_1 |r_{t-1}| + \beta_2 Q_{t-1}, \\ \nonumber
 \text{ES}_t&=& x_t Q_t, \\ \nonumber
 x_t&=& 1+\exp(\gamma_0),
\end{eqnarray}
where $\gamma_0$ is unconstrained.

\vspace{0.5cm}
{\centering
\section{\normalsize ES-CAViaR-X MODELS} \label{model_section}
}
\noindent


In this paper, the ES-CAViaR models in Taylor (2017) are extended to allow a realized measure as an explanatory variable to drive the
dynamics of risk, instead of lagged daily returns. Two general ES-CAViaR-X specifications that achieve this goal are proposed. First,
denote the realized measure on day $t$ as $X_t$, then the two specifications are:

\noindent \textbf{ES-CAViaR-AR-X:}

\begin{eqnarray} \label{re_es_caviar_ar}
Q_{t} &=& \beta_0+ \beta_1 X_{t-1} + \beta_2 Q_{t-1},\\ \nonumber
\text{ES}_t &=& Q_t-x_t, \\ \nonumber
 x_t &=&
\begin{cases}
    \gamma_0 + \gamma_1 (Q_{t-1} - r_{t-1}) + \gamma_2 x_{t-1} & \text{if } r_{t-1} \leq Q_{t-1},\\
    x_{t-1}              & \text{otherwise}.
\end{cases}
\end{eqnarray}

\noindent \textbf{ES-CAViaR-Exp-X:}

\begin{eqnarray} \label{re_es_caviar_exp}
Q_{t}&=& \beta_0+ \beta_1 X_{t-1} + \beta_2 Q_{t-1},\\ \nonumber
 \text{ES}_t &=& x_t Q_t, \\ \nonumber
 x_t &=& 1+\exp(\gamma_0),
\end{eqnarray}
where $X_t$ is a realized measure observed on day $t$. 

The AL based likelihood formation for the ES-CAViaR-X models is exactly the same as that under the framework of Taylor (2017) for the ES-CAViaR models; only the specification for the quantile (VaR) and ES series
has changed. 
It is important to note that neither the likelihood for the ES-CAViaR models nor for the ES-CAViaR-X models is a parametric likelihood or leads to a parametric MLE. The likelihood assumes a given
value for $\alpha$ during estimation, thus directly targeting a specific quantile of the conditional return distribution, without assuming it has a specific distributional form; instead the MLE
is equivalent to, and simply recasts the problem of finding, the quantile regression estimator based on minimising the quantile loss function and
the associated expected shortfall.

It is straightforward to extend the ES-CAViaR-X framework into other nonlinear forms, e.g. by choosing the nonlinear threshold quantile dynamics in Gerlach, Chen and Chan (2011). However, focus here is solely on the two ES-CAViaR-SAV type models in (\ref{re_es_caviar_ar})-(\ref{re_es_caviar_exp}).

{\centering
\section{\normalsize LIKELIHOOD AND BAYESIAN ESTIMATION} \label{beyesian_estimation_section}
\par
}
\noindent
\subsection{ES-CAViaR Likelihood Function with AL}\label{es_caviar_likelihood_section}

Exploiting the result in Koenker and Machado (1999), Gerlach, Chen and Chan (2011) employ an AL distribution combined with a CAViaR model, integrating out
the nuisance scale parameter, to allow the construction of a likelihood function and subsequently Bayesian estimation; finding favourable
properties compared to the MLE, in simulations. Taylor (2017) extended the Koenker and Machado (1999) result to
incorporate ES into the equivalent likelihood function and proposed some ES-CAViaR models to jointly estimate VaR and ES in a semi-parametric
manner. The resulting likelihood function is given in Equation (\ref{es_caviar_like_equation}).

\begin{eqnarray}\label{es_caviar_like_equation}
L(\mathbf{r};\mathbf{\theta})= \sum_{t=1}^{n} \frac{(\alpha-1)}{\text{ES}_t} \exp{\frac{(r_t-Q_t)(\alpha-I(r_t\leq Q_t))}{\alpha \text{ES}_t}}.
\end{eqnarray}

\subsection{Maximum Likelihood Estimation}

The approach in Taylor (2017, pg 5) is implemented to estimate the MLE for the ES-CAViaR-X model. Specifically, initial values for the CAViaR model component parameters are optimized separately using quantile
regression, while many initial starting value candidates are randomly sampled for the ES model component parameters; $50000$ candidate values
are considered. Each combination of initial parameter values are employed as starting values in a numerical optimisation procedure to minimise the negative of the log-likelihood function in \ref{es_caviar_like_equation}. Then, the parameter estimates that coincide with the
global minimum of this function across all starting value combinations are taken as the joint MLE vector.

Motivated by the favourable estimation results for CAViaR (Gerlach, Chen and Chan, 2011) and Conditional Autoregressive expectile (CARE) models (Gerlach and Chen, 2016), compared to the associated MLEs, a Bayesian estimator is also considered.

\subsection{Bayesian Estimation}

Given a likelihood function, and the specification of a prior distribution, Bayesian algorithms can be employed to estimate the parameters
of ES-CAViaR-AR and ES-CAViaR-Exp models. An adaptive MCMC method, adopted from that in Gerlach and Wang (2016) and Chen \emph{et al.} (2017)
is employed in this case. For the ES-CAViaR-AR model, two parameter blocks were employed in the MCMC simulation:
$\utwi{\theta_1}=(\beta_0,\beta_1,\beta_2)$, $\utwi{\theta_2}=(\gamma_0, \gamma_1, \gamma_2)$, via the motivation that parameters within
the same block are likely to be more strongly correlated in the posterior, than those between blocks, allowing faster mixing of the chain
(e.g. see Damien et al, 2013). Similarly, two blocks are also used for the ES-CAViaR-Exp model estimation: $\utwi{\theta_1}=(\beta_0,\beta_1,\beta_2)$, $\utwi{\theta_2}=(\gamma_0)$. Priors are chosen to be uninformative over the possible stationarity
and positivity regions, e.g. $\pi(\utwi{\theta})\propto I(A)$, which is a flat prior for $\utwi{\theta}$ over the region $A$.

In "burn-in" period, the "epoch" method in Chen \emph{et al.} (2017) is employed. For the initial "epoch", a Metropolis algorithm
(Metropolis \emph{et al.}, 1953) employing a mixture of 3 Gaussian proposal distributions, with a random walk mean vector, is utilised for each
block of parameters. The proposal var-cov matrix of each block in each mixture element is
$C_i \Sigma$, where $C_1 =1; C_2 =100; C_3 =0.01$ (allowing both very big, $i=2$, and very small, $i=3$ jumps), with $\Sigma$ initially set
to $\frac{2.38}{\sqrt{(d_i)}}I_{d_i}$. Here $d_i$ is the dimension of block ($i$); $I_{d_i}$ is the identity matrix of dimension $d_i$.
The covariance matrix is subsequently tuned, aiming towards a target acceptance rate of $23.4\%$ (if $d_i>4$, or $35\%$ if $2 \le d_i \le 4$,
or $44\%$ if $d_i=1$), as standard, via the algorithm of Roberts, Gelman and Gilks (1997).

In order to enhance the convergence of the chain, at the end of 1st epoch, e.g. 20,000 iterations, the covariance matrix for each parameter
block is calculated, after discarding the first e.g. 2,000 iterations, which is used in the proposal distribution in the next epoch
(of e.g. 20,000 iterations). After each epoch, the standard deviation (sd) of each parameter chain in that epoch is calculated and these
are collectively compared to the sds from the previous epoch. This process is continued until the mean absolute percentage change over the
parameters is less than a pre-specified threshold, e.g. 10\%.
In the empirical study, on average it takes 3-4 epochs to observe an absolute percentage change lower than 10\%; thus, the chains are run in
total for e.g. 60,000-80,000 iterations as a burn-in period, in the empirical parts of this paper. A final epoch is run, of say 10,000 iterates,
employing a mixture of three Gaussian proposal distributions, in an "independent" Metropolis-Hastings algorithm, in each block. The mean vector
for each block is set as the sample mean vector of the last epoch iterates (after discarding the first 2,000 iterates) for that block. The
proposal var-cov matrix in each element is $C_i \Sigma$, where $C_1 =1;C_2 =100;C_3 =0.01$ and $\Sigma$ is the sample covariance matrix of the last epoch iterates for that block (after discarding the first 2,000 iterates). This final epoch is employed as the
sample period, where all estimation, inference and forecasting is done, mainly via posterior mean estimators.

{\centering
\section{\normalsize ES-CAViaR-X SIMULATION STUDY}\label{simulation_section}
\par
}
\noindent
A simulation study is conducted to compare the properties and performance of the Bayesian method and MLE for the ES-CAViaR-X type models, with respect to parameter estimation
and one-step-ahead VaR and ES forecasting accuracy. Both the mean and Root Mean Square Error (RMSE) values are calculated over the replicated datasets for the MCMC and ML methods, to illustrate their
respective bias and precision.

1000 replicated return series are simulated from the following specific absolute value GARCH model (Taylor, 1986, Schwert, 1989) incorporating realized volatility (Abs-GARCH-X), specified as model (\ref{r_garch_x_simu}). The realized volatility data of size $n=1905$ is chosen as the 1st in-sample data of S\&P 500 forecasting study, details as in Table \ref{var_fore_table}.
The equivalent ES-CAViaR model was fit to each data set, once using MCMC and once using ML.

\begin{eqnarray} \label{r_garch_x_simu}
&&r_t= \sqrt{h_t} \epsilon_t^{*}, \\ \nonumber
&&\sqrt{h_t}= 0.02 + 0.10 \sqrt{RV_{t-1}}+ 0.85 \sqrt{h_{t-1}},  \\ \nonumber
&& \epsilon_t^{*} \stackrel{\rm i.i.d.} {\sim} N(0,1). \\ \nonumber
\end{eqnarray}

In order to calculate the corresponding ES-CAViaR true parameter values, a mapping between from the Abs-GARCH-X to the
ES-CAViaR is required. With $\text{VaR}_t=Q_t=\sqrt{h_t} \Phi^{-1}(\alpha)$, then $\sqrt{h_t} =\frac{Q_t } {\Phi^{-1}(\alpha)}= \frac{\text{VaR}_t} {\Phi^{-1}(\alpha)}$. Substituting back into the Abs-GARCH and measurement equations of model (\ref{r_garch_x_simu}),
the corresponding ES-CAViaR (without ES component) specification can be written:

\begin{eqnarray}
&& Q_{t}= 0.02 \Phi^{-1}(\alpha) + 0.10 \Phi^{-1}(\alpha) \sqrt{RV_{t-1}} + 0.85 Q_{t-1}, \\ \nonumber
\end{eqnarray}
allowing true parameter values to be calculated or read off. These true values appear in Tables \ref{simu_x_table} and \ref{simu_x_table_1}.

The situation is more complicated for the ES equation, since there is not an exact one-one mapping between the ES equation in the ES-CaViaR model and the true model in this case.
Under model (\ref{r_garch_x_simu}), the true in-sample and one-step-ahead $\alpha$ level VaR and ES forecast can be exactly calculated for each dataset; i.e. $\text{VaR}_{t}= \sqrt{h_{t}}\Phi^{-1}(\alpha)$,
and $\text{ES}_{t}= -\sqrt{h_{t}} \frac{\phi(\Phi^{-1}(\alpha))}{\alpha} $, where $\phi()$ is standard Normal pdf. For the ES-CAViaR-Exp model, the implied value of
$\gamma_0$ can be solved for each $t$; these values are then averaged and the average is treated as the true $\gamma_0$ in each simulated dataset. Finally, the average of all these "true" $\gamma_0$ values from
the 1000 datasets is presented in the "True" column of Tables \ref{simu_x_table} and \ref{simu_x_table_1}.

However, in the ES-CAViaR-AR Model, the true values of $\gamma_0, \gamma_1, \gamma_2$ cannot be solved exactly, nor in a similar manner to that for the ES-CAViaR-Exp model. A different approach is taken, where for
each dataset, $50,000$ random sets of trial values of $\gamma_0, \gamma_1, \gamma_2$ are randomly proposed, and the set that maximizes the log-likelihood (conditional upon the true series $Q_t$) is selected as the
"true" values. Finally, the average of all "true" $\gamma_0, \gamma_1, \gamma_2$ values from the 1000 datasets is presented in the "True" column Table \ref{simu_table_1} (0.1087, 0.2071 and 0.3086 respectively). The standard deviations of selected $\gamma_0, \gamma_1, \gamma_2$ across 1000 datasets are 0.0799, 0.2219 and 0.2707 respectively.

The true one-step-ahead $\alpha$ level VaR and ES forecasts can be calculated exactly as above. For each simulated dataset, the true value of $\text{VaR}_{n+1}$ and $\text{ES}_{n+1}$ are recorded;
the averages of these, over the 1000 datasets, are given in the "True" column of Table \ref{simu_x_table} and \ref{simu_x_table_1} respectively.

The ES-CAViaR-AR and ES-CAViaR-Exp models are fit to the 1000 datasets generated, once using the adaptive MCMC method and once using the ML estimator.

Estimation results for ES-CAViaR-AR are summarized in Table \ref{simu_x_table}, where boxes indicate the preferred model, in terms of minimum
bias (Mean) and maximum precision (minimum RMSE). First, both MCMC and ML generate relatively accurate parameter estimates and VaR \& ES forecasts
in this case, confirming the validity of both methods as discussed in Section \ref{beyesian_estimation_section}. However, regarding bias and
precision the results clearly favour the MCMC estimator compared to the MLE in all cases.

With respect to the ES-CAViaR-Exp model estimation, there is a reasonably consistent story. First, compared with the results from ES-CAViaR-AR model, the bias and precision results are better under ES-CAViaR-Exp. This is a result of the simpler model specification. Accurate parameter estimates
and VaR \& ES forecasting results are produced by both adaptive MCMC and ML. However,
the MCMC method produces favourable results with respect to both bias (3 out of 4) and precision (3 out of 4), and for both the VaR and ES tail
risk forecasts.

Finally, another two simulations were conducted on the performance of the Bayesian method and MLE for the ES-CAViaR type models (models (\ref{es_caviar_ar_model}) and (\ref{es_caviar_exp_model})), with details presented in Appendix \ref{simulation_return_section}.
Results marginally favoured the Bayesian approach.

\begin{table}[!ht]
\begin{center}
\caption{\label{simu_x_table} \small Summary statistics for the two estimators of the ES-CAViaR-AR-X model, with data simulated from model (\ref{r_garch_x_simu}).}\tabcolsep=10pt

\begin{tabular}{lcccccccc} \hline
$n=1905$               &             & \multicolumn{2}{c}{MCMC}      &  \multicolumn{2}{c}{ML}   \\
Parameter              &True         &Mean           &  RMSE         &Mean           & RMSE    \\ \hline
$\beta_0$         &     -0.0465 &	\fbox{-0.0579} &	\fbox{0.0568} 	&-0.0683 	& 0.0637      \\
$\beta_1$         &   -0.2326 &	\fbox{-0.2807} &	 \fbox{0.1513} & 	-0.3008 	& 0.1594      \\
$\beta_2$          &         0.8500 	& \fbox{0.8182} &	 \fbox{0.1044} 	& 0.8012& 	 0.1124 \\
$\gamma_0$          &     0.1087  	& \fbox{0.0929} 	& \fbox{0.0809} &	 0.0731 &	 0.1073     \\
$\gamma_1$           &        0.2071 	& \fbox{0.2391} 	& \fbox{0.2311} &	 0.1315 	& 0.2885 \\
$\gamma_2$            &    0.3086 	 & \fbox{0.2886} 	& \fbox{0.2844} &	 0.4702 	& 0.4728  \\
$\text{VaR}_{n+1}$     &       -1.7523 	& \fbox{-1.7424} &	 0.1236 	&-1.7179 	& \fbox{0.1232}     \\
$\text{ES}_{n+1}$     &        -2.0070 	&\fbox{-1.9396} 	 &\fbox{0.1642} &	-1.8897 	& 0.1912     \\
\hline
\end{tabular}
\end{center}
\emph{Note}:\small  A box indicates the favored estimators, based on mean and RMSE.
\end{table}

\begin{table}[!ht]
\begin{center}
\caption{\label{simu_x_table_1} \small Summary statistics for the two estimators of the ES-CAViaR-Exp-X model, with data simulated from model (\ref{r_garch_x_simu}).}\tabcolsep=10pt
\begin{tabular}{lcccccccc} \hline
$n=1905$               &             & \multicolumn{2}{c}{MCMC}      &  \multicolumn{2}{c}{ML}   \\
Parameter              &True         &Mean           &  RMSE         &Mean           & RMSE    \\ \hline
$\beta_0$         &  -0.0465  &	\fbox{-0.0593} 	 & \fbox{0.0577} 	 &-0.0656  &	 0.0647    \\
$\beta_1$         &   -0.2326 	 & \fbox{-0.2769} 	 & \fbox{0.1389}  &	-0.2874  &	 0.1441   \\
$\beta_2$          &      0.8500 	 & \fbox{0.8193}  &	 \fbox{0.0995} 	 & 0.8095 	 & 0.1063    \\
$\gamma_0$          &      -1.9264 & 	-2.1113  &	0.3414  &	\fbox{-2.0240} 	 & \fbox{0.2517}   \\
$\text{VaR}_{n+1}$     &       -1.7523  &	\fbox{-1.7457}  &	 \fbox{0.1257}  &	-1.7290  &	 0.1288   \\
$\text{ES}_{n+1}$ s     &   -2.0076 & 	\fbox{-1.9719}  &	 \fbox{0.1500} 	 &-1.9629  &	 0.1514      \\
\hline
\end{tabular}
\end{center}
\emph{Note}:\small  A box indicates the favored estimators, based on mean and RMSE.
\end{table}

{\centering
\section{\normalsize REALIZED MEASURES}\label{realized_measure_section}
\par
}
\subsection{Realized Measures}
\noindent
This section reviews the realized measures to be employed in the ES-CaViaR-X models.

For day $t$, representing the daily high, low and closing prices as $H_{t}$, $L_{t}$ and $C_{t}$, the most commonly used daily log return is:
\begin{equation}\label{return_def}
r_t= \text{log}(C_t)-\text{log}(C_{t-1}), \nonumber
\end{equation}
where $r_t^2$ is the associated (unbiased) volatility estimator. The high-low range (squared), proposed by Parkinson (1980), can be a more
efficient volatility estimator than $r_t^2$, based on the range distribution theory (see e.g. Feller, 1951):
\begin{equation}\label{range_def}
Ra_{t}^{2}=\frac {(\text{log}H_{t}-\text{log}L_{t})^2} {4\log2}, \nonumber
\end{equation}
where $4 \text{log}(2)$  scales Ra to be approximately unbiased. Several other range-based estimators, e.g.
Garman and Klass (1980); Rogers and Satchell (1991); Yang and Zhang (2000) were subsequently proposed; see Moln{\'a}r (2012) for a
full review regarding their properties.

If each day $t$ is divided into $N$ equally sized intervals of length $\Delta$, subscripted by $\Theta= {0, 1, 2, ... , N}$, several
high frequency volatility measures can be calculated. For day $t$, denote the $i$-$th$ interval closing price as $P_{t-1+i \triangle}$ and
$H_{t,i}=\text{sup}_{(i-1) \triangle<j< i \triangle}P_{t-1+j}$ and $L_{t,i}=\text{inf}_{(i-1) \triangle<j<i \triangle}{P_{t-1+j}}$ as the high and
low prices during this time interval. Then RV is proposed by Andersen and Bollerslev (1998) as:
\begin{equation}\label{rv_def2}
RV_{t}^{\triangle}=\sum_{i=1}^{N} [log(P_{t-1+i \triangle})-log(P_{t-1+(i-1)\triangle})]^{2}
\end{equation}
Martens and van Dijk (2007) and Christensen and Podolskij (2007) developed the Realized Range, which sums the squared intra-period
ranges:
\begin{equation}\label{rrv_def}
RR_{t}^{\triangle}= \frac {\sum_{i=1}^{N}(\text{log}H_{t,i}-\text{log}L_{t,i})^2}{4\log2}.
\end{equation}

If each day $t$ is divided into $N$ equally sized intervals of length $\Delta$, subscripted by $\Theta= {0, 1, 2, ... , N}$, several
high frequency volatility measures can be calculated. For day $t$, denote the $i$-$th$ interval closing price as $P_{t-1+i \triangle}$ and
$H_{t,i}=\text{sup}_{(i-1) \triangle<j< i \triangle}P_{t-1+j}$ and $L_{t,i}=\text{inf}_{(i-1) \triangle<j<i \triangle}{P_{t-1+j}}$ as the high and
low prices during this time interval. Then RV is proposed by Andersen and Bollerslev (1998) as:
\begin{equation}\label{rv_def2}
RV_{t}^{\triangle}=\sum_{i=1}^{N} [log(P_{t-1+i \triangle})-log(P_{t-1+(i-1)\triangle})]^{2}.
\end{equation}
Martens and van Dijk (2007) and Christensen and Podolskij (2007) developed the Realized Range, which sums the squared intra-period
ranges:
\begin{equation}\label{rrv_def}
RR_{t}^{\triangle}= \frac {\sum_{i=1}^{N}(\text{log}H_{t,i}-\text{log}L_{t,i})^2}{4\log2}
\end{equation}

Through theoretical derivation and simulation, Martijns and van Dijk (2007) show that RR is a competitive, and sometimes more efficient,
volatility estimator than RV under some micro-structure conditions and levels. Gerlach and Wang (2016) confirm that RR can provide extra
efficiency in empirical tail risk forecasting, when employed as the measurement equation variable in an Re-GARCH model.
To further reduce the effect
of microstructure noise, Martens and van Dijk (2007) presented a scaling process, as in Equations (\ref{rv_scale}) and (\ref{rrv_scale}).
\begin{eqnarray}\label{rv_scale}
ScRV_{t}^{\triangle}= \frac {\sum_{l=1}^{q}RV_{t-l}}{\sum_{l=1}^{q}RV_{t-l}^{\triangle}}RV_{t}^{\triangle},
\end{eqnarray}
\begin{eqnarray}\label{rrv_scale}
ScRR_{t}^{\triangle}= \frac {\sum_{l=1}^{q}RR_{t-l}}{\sum_{l=1}^{q}RR_{t-l}^{\triangle}}RR_{t}^{\triangle},
\end{eqnarray}
\noindent
where $RV_{t}$ and $RR_{t}$ represent the daily squared return and squared range on day $t$,, respectively. This scaling process is
inspired by the fact that the daily squared return and range are each less affected by micro-structure noise than their high frequency
counterparts, thus can be used to scale and smooth RV and RR, creating less micro-structure sensitive measures.

Zhang, Mykland and A\"{i}t-Sahalia (2005) proposed a sub-sampled process to further smooth out micro-structure noise.
For day $t$, $N$ equally sized samples are grouped into $M$ non-overlapping subsets $X^{(m)}$ with size $N/M=n_{k}$, which means:
\begin{equation}
X = \bigcup_{m=1}^{M} X^{(m)}, \; \mbox{where} \; X^{(k)}  \cap X^{(l)} = \emptyset, \;  \mbox{when}  \;  k \neq l.  \nonumber
\end{equation}

Then sub-sampling is implemented on the subsets $X^{(i)}$ with $n_{k}$ interval:
\begin{equation}
X^{(i)}= {i, i+n_k,...,i+n_k(M-2), i+n_k(M-1)}, \; \mbox{where} \;  i= {0,1,2...,n_k-1}.  \nonumber
\end{equation}

Representing the log closing price at the $i$-$th$ interval of day $t$ as $C_{t,i}=P_{t-1+i\triangle}$, the RV with subsets $X^{i}$ is:
\begin{equation}
RV_{t,i}= \sum_{m=1}^{M} (C_{t,i+n_{k}m}-C_{t,i+n_{k}(m-1)})^{2}; \; \mbox{where} \; i= {0,1,2...,n_k-1}.  \nonumber
\end{equation}

We have $T/M$ RV with $T/N$ sub-sampling for day $t$ as (supposing there are $T$ minutes per trading day):

\begin{equation}\label{ssrv_def_2}
SSRV_{t,T/M,T/N}^{\triangle}= \frac{\sum_{i=0}^{n_k-1} RV_{t,i} } {n_k},
\end{equation}

then, denoting high and low prices during the interval $i+n_{k}(m-1)$ and $i+n_{k}m$ as
$H_{t,i}=\sup_{(i+n_{k}(m-1))\triangle<j<(i+n_{k}m) \triangle}P_{t-1+j}$ and
$L_{t,i}=\inf_{(i+n_{k}(m-1))\triangle<j<(i+n_{k}m) \triangle}{P_{t-1+j}}$ respectively, we propose the $T/M$ RR with $T/N$ sub-sampling as:
\begin{equation}
RR_{t,i}= \sum_{m=1}^{M} (H_{t,i}-L_{t,i})^2 ; \; \mbox{where} \; i= {0,1,2...,n_k-1}.
\end{equation}
\begin{equation}\label{ssrr_def_2}
SSRR_{t,T/M,T/N}^{\triangle}= \frac{\sum_{i=0}^{n_k-1} RR_{t,i} } { 4 \mbox{log}2 n_k}.
\end{equation}

For example, the 5 mins RV and RR with 1 min sub-sampling for any day can be calculated, respectively, as below :
\begin{eqnarray}  \nonumber
&&RV_{5,1,0}=(\log C_{t5}-\log C_{t0})^2+(\log C_{t10}-\log C_{t5})^2+... \\  \nonumber
&&RV_{5,1,1}=(\log C_{t6}-\log C_{t1})^2+(\log C_{t11}-\log C_{t6})^2+... \\ \nonumber
&&\vdots \\ \nonumber
&&RV_{5,1,4}=(\log C_{t9}-\log C_{t4})^2+(\log C_{t14}-\log C_{t9})^2+... \\ \nonumber
&&SSRV_{5,1}^{\triangle}=\frac{\sum_{i=0}^{4}RV_{5,1,i}} {5}.  \nonumber
\end{eqnarray}

\begin{eqnarray}  \nonumber
&&RR_{5,1,0}=(\log H_{t0 < t < t5}-\log L_{t0 < t <  t5})^2+(\log H_{t5 < t < t10}-\log L_{t5 < t <  t10})^2+... \\ \nonumber
&&RR_{5,1,1}=(\log H_{t1< t < t6}-\log L_{t1 < t <  t6})^2+(\log H_{t6 < t < t11}-\log L_{t6 < t <  t11})^2+... \\ \nonumber
&&\vdots \\ \nonumber
&&RR_{5,1,4}=(\log H_{t4< t < t9}-\log L_{t4 < t <  t9})^2+(\log H_{t9 < t < t14}-\log L_{t9 < t <  t14})^2+... \\ \nonumber
&&SSRR_{5,1}^{\triangle}=\frac{\sum_{i=0}^{4}RR_{5,1,i}} {4 \log (2)5}. \nonumber
\end{eqnarray}
Only intra-day return and range on the 5 minute frequency, additionally with 1 minute sub-sampling when employed, are considered in this paper.

{\centering
\section{\normalsize DATA and EMPIRICAL STUDY}\label{data_empirical_section}
\par
}
\subsection{Data Description}
Daily and high frequency data, observed at 1-minute and 5-minute frequency, including open, high, low and closing prices, are downloaded from
Thomson Reuters Tick History. Data are collected for 7 market indices: S\&P500, NASDAQ (both US), Hang Seng (Hong Kong), FTSE 100 (UK),
DAX (Germany), SMI (Swiss) and ASX200 (Australia); as well as for 2 individual assets: IBM and GE (both US); the time period is Jan 2000 to June 2016, however
for GE the data starts in May 2000, after a $3:1$ stock split in April, 2000.

\begin{figure}[htp]
     \centering
\includegraphics[width=.9\textwidth]{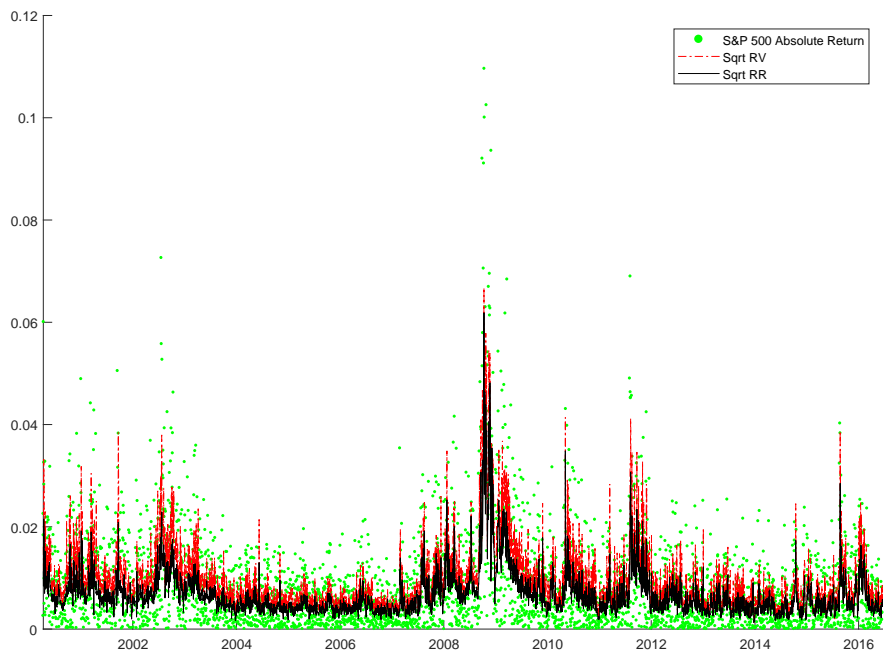}
\caption{\label{Fig2} S\&P 500 absolute return, $\sqrt{RV}$ and $\sqrt{RR}$ Plots.}
\end{figure}

The daily return, daily range, daily range plus overnight price jump and the daily RV and RR measures are calculated using 5 minute data; 1-minute data are employed to produce daily scaled and
sub-sampled RV and RR measures; $q=66$ is employed for the scaling process, i.e. around 3 months.
Figure \ref{Fig2} plots the S\&P 500 absolute value of daily returns, as well as $\sqrt{RV}$ and $\sqrt{RR}$ for expositional purposes.

\subsection{Tail Risk Forecasting}

Both daily Value-at-Risk (VaR) and Expected Shortfall (ES) are forecast one day ahead for the 7 indices and the 2 asset series, using
$\alpha=0.01$, as recommended in the Basel II and III Capital Accord.

A rolling window with fixed in-sample size is employed for estimation to produce each 1 step ahead forecast in the forecast period; the
in-sample size $n$ is given in Table \ref{var_fore_table} for each series, which differs due to non-trading days in each market. In order to see
the performance during the GFC period, the initial date of the forecast sample is chosen as the beginning of 2008. On average,
2111 one day ahead VaR and ES forecasts are generated for each return series from a range of models. These include the proposed ES-CAViaR-X
type models (estimated via MCMC) with different input realized measures: RV \& RR, scaled RV \& RR and
sub-sampled RV \& RR. The original ES-CAViaR models (estimated with adaptive MCMC) of Taylor (2017) are also included in the study.
The conventional GARCH, EGARCH and GJR-GARCH with Student-t distribution, CARE-SAV and Re-GARCH with Gaussian and Student-t observation
equation error distributions, are also included, for the purpose of comparison. Further, a filtered GARCH (GARCH-HS) approach is also included,
where a GARCH-t is fit to the in-sample data, then standardised VaR and ES are estimated via historical simulation, using all in-sample data, i.e. $r_1,\ldots,r_n$ divided by their GARCH-estimated conditional standard deviation (i.e. $r_t/\sqrt{\hat{h_t}}$).
Then final forecasts of VaR, ES are found by multiplying the standardised VaR, ES estimates by the forecast $\sqrt{\hat{h}_{n+1}}$
from the GARCH-t model. These reference, comparative models are estimated by ML, using the Econometrics toolbox in Matlab (GARCH-t, EGARCH-t, GJR-t
and GARCH-HS) or code developed by the authors (CARE-SAV and Re-GARCH). The actual forecast sample sizes $m$, in each series, are given in Table \ref{var_fore_table}.

The VaR violation rate (VRate) is employed to initially assess the VaR forecasting accuracy: simply the proportion of returns
that exceed the forecasted VaR level in the forecast period.
Models with VRate closer to nominal quantile level $\alpha=0.01$ are preferred.


Having a VRate close to $\alpha$ is a necessary but not sufficient condition to guarantee an accurate forecasting model.
Thus several standard quantile accuracy and independence tests are also employed: the unconditional coverage (UC) and conditional
coverage (CC) tests of Kupiec (1995) and Christoffersen (1998) respectively, as well as the dynamic quantile (DQ) test of
Engle and Manganelli (2004) and the VQR test of Gaglione \emph{et al.} (2011). Finally, the standard quantile loss function is also employed to compare
the models for VaR forecast accuracy. Quantiles are elicitable, in the sense defined by Gneiting (2012), since the standard quantile loss function
is strictly consistent, i.e. the expected loss is a minimum at the true quantile series. Thus, the most accurate VaR forecasting model
should minimise the quantile loss function, given as:
\begin{equation}\label{q_loss}
\sum_{t=n+1}^{n+m}(\alpha-I(r_t<Q_t))(r_t-Q_t)  \,\, ,
\end{equation}
where $Q_{n+1},\ldots,Q_{n+m}$ is a series of quantile forecasts at level $\alpha$ for the observations $r_{n+1},\ldots,r_{n+m}$.

\subsubsection{\normalsize Value at Risk}

Table \ref{var_fore_table} presents the VRates for each model over the 9 return series,
while Table \ref{Summ_var_fore} summarizes those results; $\alpha = 0.01$ or 1\% is the target rate. A box indicates the model with VRate
closest to 1\% in each market, while bolding indicates the VRate is significantly different to 1\% by the UC test. 


In 5 out of 9 series one of the ES-CAViaR-X models has VRate closest to nominal. From Table \ref{Summ_var_fore}, the Gt-HS has average VRate closest to 1\% quantile level followed closely by the the ES-CAViaR-X model
employing SSRR, while the ES-CAViaR-X models employing SSRV and SSRR have median VRate closest to nominal. 
All the models were anti-conservative, as expected since the forecast sample includes the GFC period, having VRates on average (and median) above 1\%: the Re-GARCH-GG was most anti-conservative, generating 80-90\% too many violations, not surprising since it is the only model employing the Gaussian error distribution.


\begin{table}[!ht]
\begin{center}
\caption{\label{var_fore_table} \small 1\% VaR Forecasting VRate with different models on 7 indices and 2 assets.}\tabcolsep=10pt
\tiny
\begin{tabular}{lccccccccccc} \hline
Model               &  S\&P 500          &NASDAQ             &HK              &FTSE            &DAX              &SMI              &ASX200           &IBM            &GE    \\ \hline
G-t                 &\bf{1.467\%}&\bf{1.895\%}&\bf{1.652\%}&\bf{1.731\%}&1.362\%&\bf{1.617\%}&\bf{1.702\%}&1.183\%&\cb{0.945\%}\\
EG-t                &\bf{1.514\%}&\bf{1.611\%}&1.215\%&\bf{1.777\%}&1.408\%&\bf{1.712\%}&\bf{1.466\%}&1.183\%&1.040\%\\
GJR-t               &\bf{1.467\%}&\bf{1.563\%}&1.263\%&\bf{1.777\%}&1.408\%&\bf{1.759\%}&\bf{1.513\%}&\cb{1.088\%}&1.040\%\\
Gt-HS               &1.230\%&\bf{1.563\%}&1.263\%&1.123\%&\cb{1.127\%}&1.284\%&0.898\%&\fbox{1.041\%}&1.181\%\\
CARE                &1.278\%&\bf{1.563\%}&\cb{1.020\%}&1.310\%&1.221\%&1.284\%&1.229\%&1.183\%&1.371\%\\
RG-RV-GG            &\bf{2.130\%}&\bf{1.942\%}&\bf{2.818\%}&\bf{1.777\%}&\bf{2.300\%}&\bf{1.807\%}&\bf{1.560\%}&1.419\%&1.323\%\\
RG-RV-tG            &\bf{1.467\%}&\fbox{1.326\%}&\bf{1.992\%}&1.310\%&\bf{1.596\%}&\fbox{1.141\%}&1.229\%&0.851\%&0.803\%\\
ES-CAV-AR           &\bf{1.467\%}&\bf{1.516\%}&1.215\%&1.216\%&1.268\%&\cb{1.236\%}&\fbox{0.946\%}&1.230\%&\fbox{0.992\%}\\
ES-CAV-AR-RV      &1.372\%&\bf{1.658\%}&\bf{2.138\%}&1.076\%&1.174\%&1.379\%&1.040\%&1.183\%&1.134\%\\
ES-CAV-AR-RR      &\cb{1.230\%}&\bf{1.516\%}&1.215\%&\fbox{0.935\%}&1.268\%&\bf{1.522\%}&0.851\%&1.230\%&\bf{1.560\%}\\
ES-CAV-AR-ScRV    &1.420\%&\bf{1.658\%}&1.166\%&1.216\%&\fbox{1.080\%}&\cb{1.236\%}&\cb{1.040\%}&\cb{1.088\%}&1.229\%\\
ES-CAV-AR-ScRR    &1.420\%&\bf{1.942\%}&\fbox{0.972\%}&1.123\%&1.315\%&1.284\%&0.804\%&1.135\%&\bf{1.465\%}\\
ES-CAV-AR-SSRV    &1.420\%&\bf{1.705\%}&1.166\%&\cb{1.029\%}&1.408\%&1.427\%&0.851\%&1.135\%&1.134\%\\
ES-CAV-AR-SSRR    &\fbox{1.136\%}&\bf{1.516\%}&1.166\%&\fbox{0.935\%}&1.221\%&\bf{1.522\%}&0.804\%&1.230\%&\bf{1.512\%}\\
ES-CAV-Exp          &1.278\%&\cb{1.421\%}&1.166\%&1.216\%&1.315\%&\cb{1.236\%}&\fbox{0.946\%}&1.277\%&\cb{0.945\%}\\
ES-CAV-Exp-RV     &1.325\%&\bf{1.753\%}&\bf{2.089\%}&1.123\%&1.315\%&1.284\%&\fbox{0.946\%}&1.135\%&1.181\%\\
ES-CAV-Exp-RR     &1.183\%&\bf{1.468\%}&1.166\%&0.935\%&1.221\%&1.427\%&0.757\%&1.230\%&\bf{1.560\%}\\
ES-CAV-Exp-ScRV   &\bf{1.562\%}&\bf{1.705\%}&1.166\%&1.216\%&1.174\%&1.284\%&\cb{1.040\%}&\cb{1.088\%}&\bf{1.512\%}\\
ES-CAV-Exp-ScRR   &1.420\%&\bf{1.800\%}&\cb{1.020\%}&1.169\%&1.268\%&1.331\%&0.851\%&\cb{1.088\%}&\bf{1.465\%}\\
ES-CAV-Exp-SSRV   &1.372\%&\bf{1.658\%}&1.215\%&\fbox{0.935\%}&1.408\%&1.331\%&0.804\%&1.183\%&1.418\%\\
ES-CAV-Exp-SSRR   &\fbox{1.136\%}&\bf{1.516\%}&1.166\%&0.842\%&1.268\%&1.379\%&0.709\%&1.135\%&\bf{1.607\%}\\ \hline
%
m                   &2113&2111&2058&2138&2130&2103&2115&2114&2116\\
n                   &1905&1892&1890&1943&1936&1930&1871&1916&1839\\
\hline
\end{tabular}
\end{center}
\emph{Note}:\small  Box indicates the favored models based on VRate, blue shading indicates the 2nd ranked model, in each series, whilst bold indicates the violation rate is
significantly different to 1\% by the UC test. $m$ is the out-of-sample size, and $n$ is in-sample size. RG stands for the Realized-GARCH
type models, and ES-CAV-AR-RV, etc, represent the ES-CAViaR-X type models. 
\end{table}

\begin{table}[!ht]
\begin{center}
\caption{\label{Summ_var_fore} \small Summary of 1\% VaR Forecast VRates, for different models on 7 indices and 2 assets.}\tabcolsep=10pt
\begin{tabular}{lccc} \hline
Model    &Mean              &Median          \\ \hline 
G-t&\cred{1.505\%}& \cred{1.608\%}\\
EG-t&1.437\%&1.466\%\\
GJR-t    &1.432\%&1.466\%\\
Gt-HS    &\fbox{1.190\%}&1.183\%\\
CARE          &1.274\%&1.277\%\\
RG-RV-GG   &\bf{1.895\%}&\bf{1.798\%}\\
RG-RV-tG      &1.300\%&1.325\%\\
ES-CAV-AR   &1.232\%&1.230\%\\
ES-CAV-AR-RV&1.348\%&1.183\%\\
ES-CAV-AR-RR&1.258\%&\cb{1.230\%}\\
ES-CAV-AR-ScRV&1.237\%&\cb{1.230\%}\\
ES-CAV-AR-ScRR&1.274\%&1.277\%\\
ES-CAV-AR-SSRV&1.253\%&\fbox{1.135\%}\\
ES-CAV-AR-SSRR&1.226\%&\cb{1.230\%}\\
ES-CAV-Exp      &1.200\%&\cb{1.230\%}\\
ES-CAV-Exp-RV&1.348\%&1.277\%\\
ES-CAV-Exp-RR&1.216\%&\cb{1.230\%}\\
ES-CAV-Exp-ScRV&1.305\%&\cb{1.230\%}\\
ES-CAV-Exp-ScRR&1.269\%&1.277\%\\
ES-CAV-Exp-SSRV&1.258\%&1.325\%\\
ES-CAV-Exp-SSRR&\cb{1.195\%}&\fbox{1.135\%}\\ \hline
m&2110.88&2114\\
n&1902.44&1905\\
\hline
\end{tabular}
\end{center}
\emph{Note}:\small  Box indicates the favoured model, blue shading indicates the 2nd ranked model, bold indicates the least favoured model,
red shading indicates the 2nd lowest ranked model, in each column.
\end{table}

The quantile loss results are presented in Table \ref{quanitl_loss_table} and \ref{quanitl_loss_summ_table}, showing the quantile loss values
for each model for each series. Again 5 of the 9 series have the lowest loss given by one of the ES-CAViaR-X models. The lowest average loss, and average rank loss, respectively, across the series are given by ES-CAViaR-X models employing
SSRV and SSRR respectively; shown by boxes in Table \ref{quanitl_loss_summ_table}. The quantile loss values for the original ES-CAViaR type models are typically much higher than those for the ES-CAViaR-X models
and typically closer to that for the EGARCH-t, GJR-t, Gt-HS, and CARE models, which in general produce relatively higher quantile loss values.

\begin{table}[!ht]
\begin{center}
\caption{\label{quanitl_loss_table} \small 1\% VaR Forecasting quantile loss on 7 indices and 2 assets.}\tabcolsep=10pt
\tiny
\begin{tabular}{lccccccccccc} \hline
Model    &S\&P 500          &NASDAQ             &HK              &FTSE            &DAX              &SMI              &ASX200           &IBM            &GE    \\ \hline
G-t&81.84&92.12&98.40&81.46&\cred{93.36}&\cred{88.03}&69.66&114.77&128.83\\
EG-t&80.28&92.20&\fbox{90.28}&76.90&92.22&83.06&67.32&\cred{115.01}&127.48\\
GJR-t    &77.62&89.80&92.21&77.91&93.95&85.67&67.88&116.00&126.55\\
Gt-HS    &81.77&91.49&96.91&80.25&93.90&86.26&69.54&\bf{116.25}&130.58\\
CARE          &\bf{84.22}&\bf{95.53}&92.95&\bf{82.69}&93.32&\bf{89.77}&\bf{77.28}&111.99&\bf{142.70}\\
RG-RV-GG   &80.03&87.26&\bf{119.03}&78.12&\bf{95.23}&83.4&66.14&109.19&112.52\\
RG-RV-tG      &77.12&\fbox{85.31}&\cred{108.58}&77.02&91.71&82.04&\fbox{65.45}&\fbox{108.00}&112.48\\
ES-CAV-AR&\cred{84.19}&\cred{93.51}&94.85&81.09&92.81&86.28&\cred{71.90}&111.27&\cred{133.50}\\
ES-CAV-AR-RV&75.24&90.13&105.03&77.95&91.43&80.58&65.83&112.19&112.38\\
ES-CAV-AR-RR&73.70&87.39&99.85&77.87&90.47&\fbox{78.05}&67.87&112.91&\fbox{110.90}\\
ES-CAV-AR-ScRV&75.90&90.32&96.26&77.04&92.94&81.95&66.74&110.98&114.28\\
ES-CAV-AR-ScRR&75.36&90.02&\cb{90.72}&77.51&91.29&78.36&68.36&109.25&114.20\\
ES-CAV-AR-SSRV&74.3&88.68&91.21&\cb{75.81}&90.48&78.55&66.40&109.81&115.08\\
ES-CAV-AR-SSRR&73.64&\cb{86.80}&95.60&76.48&\fbox{89.90}&78.38&66.77&110.17&116.22\\
ES-CAV-Exp&83.46&93.25&95.31&\cred{81.65}&93.19&85.69&71.72&110.24&132.14\\
ES-CAV-Exp-RV&75.14&90.61&105.01&77.66&91.64&80.57&\cb{65.58}&111.68&112.44\\
ES-CAV-Exp-RR&\cb{73.21}&87.89&99.69&77.67&90.25&78.15&67.58&111.55&\cb{111.26}\\
ES-CAV-Exp-ScRV&76.12&90.46&96.37&77.1&93.22&82.22&66.61&110.24&117.71\\
ES-CAV-Exp-ScRR&75.36&90.94&91.11&77.38&91.17&78.50&68.21&\cb{108.88}&114.15\\
ES-CAV-Exp-SSRV&74.00&88.85&91.3&\fbox{75.79}&90.35&78.59&66.23&109.68&116.96\\
ES-CAV-Exp-SSRR&\fbox{73.16}&87.45&96.13&76.25&\cb{90.08}&\cb{78.35}&66.47&110.81&118.21\\
\hline
\end{tabular}
\end{center}
\emph{Note}:\small  Box indicates the favoured model, blue shading indicates the 2nd ranked model, bold indicates the least favoured model,
red shading indicates the 2nd lowest ranked model, in each column.
\end{table}


\begin{table}[!ht]
\begin{center}
\caption{\label{quanitl_loss_summ_table} \small  Quantile loss function summary for different models on 7 indices and 2 assets.}\tabcolsep=10pt
\begin{tabular}{lccc} \hline
Model    &Mean              &Mean rank          \\ \hline
G-t&94.27&\cred{17.78}\\
EG-t&91.64&12.44\\
GJR-t    &91.95&14.22\\
Gt-HS    &94.11&17.44\\
CARE          &\bf{96.72}&\bf{18.33}\\
RG-RV-GG   &92.33&11.56\\
RG-RV-tG      &89.75&7.78\\
ES-CAV-AR&\cred{94.38}&16.67\\
ES-CAV-AR-RV&90.08&10.33\\
ES-CAV-AR-RR&88.78&8.33\\
ES-CAV-AR-ScRV&89.6&10.67\\
ES-CAV-AR-ScRR&88.36&8.00\\
ES-CAV-AR-SSRV&\fbox{87.81}&\cb{6.00}\\
ES-CAV-AR-SSRR&\cb{88.22}&\fbox{5.89}\\
ES-CAV-Exp&94.07&16.11\\
ES-CAV-Exp-RV&90.04&9.89\\
ES-CAV-Exp-RR&88.58&7.56\\
ES-CAV-Exp-ScRV&90.00&11.67\\
ES-CAV-Exp-ScRR&88.41&8.22\\
ES-CAV-Exp-SSRV&87.97&\fbox{5.89}\\
ES-CAV-Exp-SSRR&88.55&6.22\\
\hline
\end{tabular}
\end{center}
\emph{Note}:\small  Boxes indicate the favoured model, blue shading indicates the 2nd ranked model, bold indicates the least favoured model,
red shading indicates the 2nd lowest ranked model, in each column. "Mean rank" is the average rank across the 7 markets and 2 assets for the
quantile loss function, over the 21 models: lower is better.
\end{table}

\begin{figure}[htp]
     \centering
\includegraphics[width=.9\textwidth]{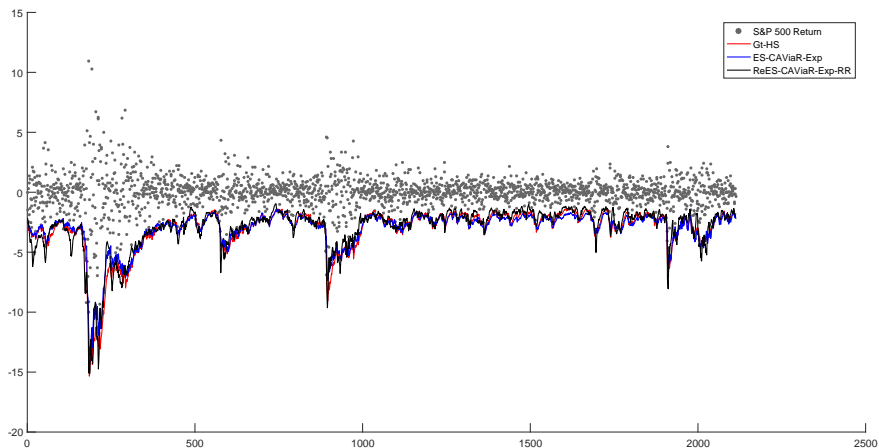}
\caption{\label{var_forecast_fig} S\&P 500 VaR Forecasts with Gt-HS, ES-CAViaR-Exp and ES-CAViaR-Exp-RR.VRates: 1.230\%, 1.278\% and 1.183\%.}
\end{figure}

\begin{figure}[htp]
     \centering
\includegraphics[width=.9\textwidth]{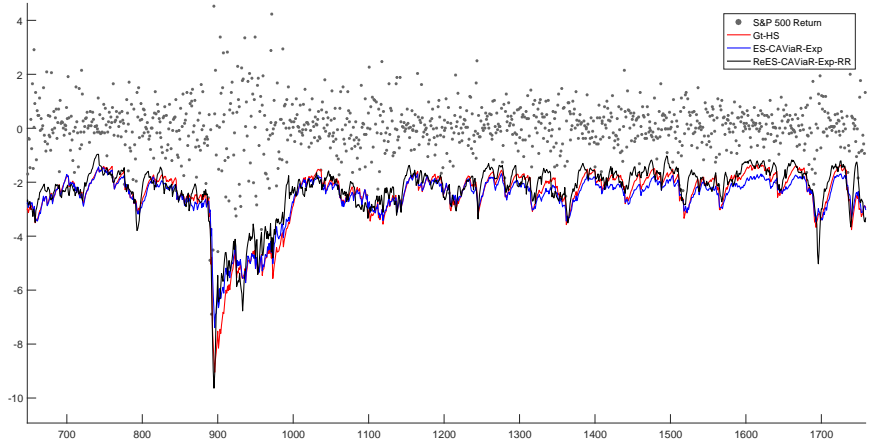}
\caption{\label{var_forecast_fig1} S\&P 500 VaR Forecasts with Gt-HS, ES-CAViaR-Exp and ES-CAViaR-Exp-RR.
VRates: 1.230\%, 1.278\% and 1.183\%. Quantile loss: 81.77,  83.46,  73.21}
\end{figure}

Figure \ref{var_forecast_fig} and \ref{var_forecast_fig1} provide evidence on how and why the proposed ES-CAViaR-X type models generate clearly
lower quantile loss compared with other models, combined with relatively accurate VRates. Specifically, the VaR violation rates for the Gt-HS, ES-CAViaR-Exp and ES-CAViaR-Exp-RR models are 1.230\%, 1.278\% and 1.183\% respectively, for the S\&P500 returns, i.e. the three
models are very similar by that metric. However, from Table \ref{quanitl_loss_table}, the quantile loss values for the 3 models are 81.77,  83.46
and 73.21 respectively, meaning the ES-CAViaR-Exp-RR model is the most accurate model having clearly the lowest quantile loss. Through close inspection of Figure \ref{var_forecast_fig1}, the ES-CAViaR-Exp and Gt-HS have VaR forecasts typically quite close together in value, driving their very close quantile loss values. However, both these models generate clearly more extreme (in the negative direction) VaR forecasts
on most days in the US market, compared to the ES-CAViaR-Exp-RR. This means that the capital set aside by financial institutions to cover extreme losses, based on such VaR forecasts, is usually at a higher level for the Gt-HS or ES-CAViaR-Exp models, than for the ES-CAViaR-Exp-RR.

In other words, the ES-CAViaR-Exp-RR model produces VaR forecasts that are relatively close to nominal VRate, are closer to the true VaR series, as measured by the loss function, and are closer to the data and less extreme, implying that lower amounts of capital are needed to protect against market risk. Given the forecasting steps  $m=2113$ for S\&P 500, the forecasts from ES-CAViaR-Exp-RR were less extreme than those from ES-CAViaR-Exp on 1443 days (68\%) in the forecast period. This suggests a higher level of information (and cost) efficiency regarding risk levels for the ES-CAViaR-Exp-X model, likely coming from the increased statistical efficiency of the realized range series over squared returns, compared to the ES-CAViaR-Exp and Gt-HS models. Since the economic capital is determined by financial institutions' own model and should be
directly proportional to the VaR forecast, the ES-CAViaR-Exp-RR model is able to decrease the cost capital allocation and increase the
profitability of these institutions, by freeing up part of the regulatory capital from risk coverage into investment, while still
providing sufficient and more than adequate protection against violations. The more accurate and often less extreme VaR forecasts produced
by ES-CAViaR-Exp-RR are particularly strategically important to the decision makers in the financial sector. This extra efficiency is also often observed for the ES-CAViaR-AR-X and ES-CAViaR-Exp-X type models in the other markets/assets.

Further, during the periods with high volatility including GFC, when there is a persistence of extreme returns, the ES-CAViaR-Exp-RR VaR forecasts "recover" the fastest among the 3 models, presented through close inspection as in Figure \ref{var_forecast_fig1}, in terms of being marginally the fastest to produce forecasts that again rejoin and follow the tail of the return data. Traditional GARCH models tend to over-react to extreme events and to be subsequently very slow to recover, due to their oft-estimated very high level of persistence, as discussed in Harvey and Chakravarty (2009). ES-CAViaR-X models clearly improve the performance on this aspect. Generally, the ES-CAViaR-X type models better describe the dynamics in the volatility, compared to the traditional GARCH model and original ES-CAViaR type models, thus largely improving the responsiveness and accuracy of the risk level forecasts, especially after high volatility periods.

Several tests are employed to statistically assess the forecast accuracy and independence of violations from each VaR forecast model.
Table \ref{var_backtest_table} shows the number of return series (out of 9) in which each 1\% VaR forecast model is rejected for each test,
conducted at a 5\% significance level. The ES-CAViaR-X type models are generally less likely to be rejected by the back tests
compared to other models, and the RG-RV-tG, ES-CAViaR-AR-ScRV, ES-CAViaR-Exp-RV and ES-CAViaR-Exp-SSRR achieved the least number of rejections (3).
The G-t is rejected in all 9 series, and the EG-t and Re-GARCH-GG models are rejected in 8 series, respectively.

\begin{table}[!ht]
\begin{center}
\caption{\label{var_backtest_table} \small Counts of 1\% VaR  rejections with UC, CC, DQ and VQR tests for different models on 7 indices and 2 assets.}\tabcolsep=10pt
\footnotesize
\begin{tabular}{lcccccc} \hline
Model    &  UC & CC & DQ1 & DQ4 & VQR  & Total \\ \hline
G-t&6&6&7&7&5& \bf{9}\\
EG-t&5&3&4&7&2&\cred{8}\\
GJR-t    &5&3&6&5&3&7\\
Gt-HS    &1&1&1&3&1&\cb{4}\\
CARE          &1&1&0&5&0&5\\
RG-RV-GG   &7&7&7&7&5&\cred{8}\\
RG-RV-tG      &3&2&2&1&3&\fbox{3}\\
ES-CAV-AR&2&0&0&4&1&6\\
ES-CAV-AR-RV&2&2&2&2&3&\cb{4}\\
ES-CAV-AR-RR&3&2&2&2&2&5\\
ES-CAV-AR-ScRV&1&2&2&3&1&\fbox{3}\\
ES-CAV-AR-ScRR&2&2&3&4&2&6\\
ES-CAV-AR-SSRV&1&1&1&3&2&\cb{4}\\
ES-CAV-AR-SSRR&3&1&1&3&2&6\\
ES-CAV-Exp&0&0&0&4&0&\cb{4}\\
ES-CAV-Exp-RV&2&2&2&3&2&\fbox{3}\\
ES-CAV-Exp-RR&2&3&3&3&2&5\\
ES-CAV-Exp-ScRV&3&2&2&3&1&\cb{4}\\
ES-CAV-Exp-ScRR&2&2&3&4&1&5\\
ES-CAV-Exp-SSRV&1&1&2&5&2&6\\
ES-CAV-Exp-SSRR&2&2&2&2&1&\fbox{3}\\
\hline
\end{tabular}
\end{center}
\emph{Note}:\small Box indicates the model with least number of rejections, blue shading indicates the model with 2nd least number of rejections, bold indicates the model with the highest number of rejections, red shading indicates the model 2nd highest number of rejections.
All tests are conducted at 5\% significance level.
\end{table}

\subsubsection{\normalsize Expected Shortfall}
The same set of 21 models are employed to generate 1-step-ahead forecasts of 1\% ES for all 9 series during the forecast sample periods.

\begin{figure}[htp]
     \centering
\includegraphics[width=.9\textwidth]{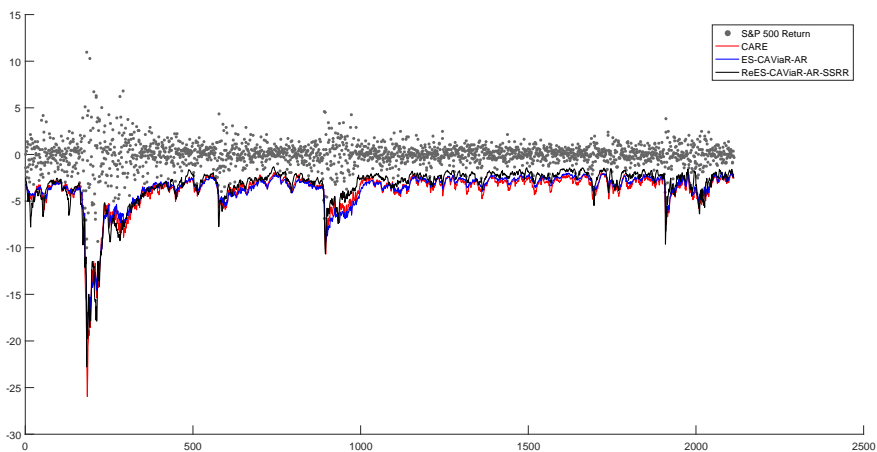}
\caption{\label{Fig_es_fore} S\&P 500 ES Forecasts with CARE, ES-CAViaR-AR and ES-CAViaR-AR-SSRR. ESRates: 0.284\%, 0.237\% and 0.426\%.}
\end{figure}

\begin{figure}[htp]
     \centering
\includegraphics[width=.9\textwidth]{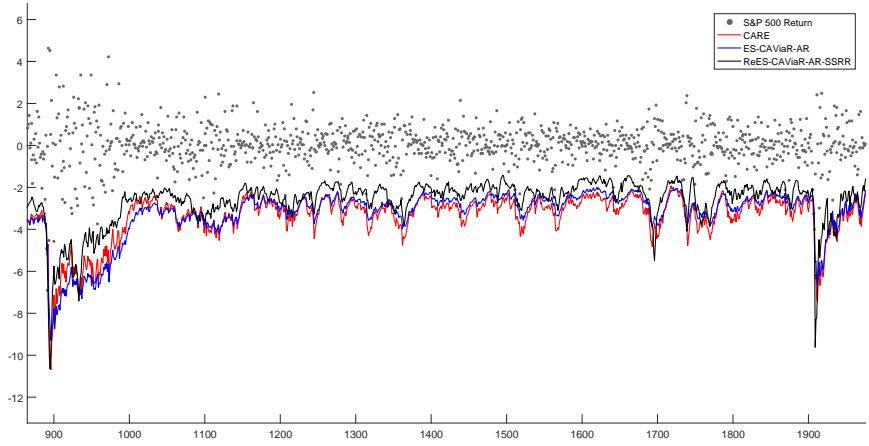}
\caption{\label{Fig_es_fore_zoom_in} S\&P 500 ES Forecasts with CARE, ES-CAViaR-AR and ES-CAViaR-AR-SSRR. ESRates: 0.284\%, 0.237\% and 0.426\%. VaR\&ES join loss:   4836.65,  4844.03,  4499.02.}
\end{figure}

First, to demonstrate the extra forecasting efficiently can be gained by employing the proposed ES-CAViaR-X type models,
Figure \ref{Fig_es_fore} and \ref{Fig_es_fore_zoom_in} visualize the ES forecasts from CARE, ES-CAViaR-AR and ES-CAViaR-AR-SSRR. Specifically, the
ES violation rate of CARE, ES-CAViaR-AR and ES-CAViaR-AR-SSRR models are 0.284\%, 0.237\% and 0.426\% respectively,
for S\&P500. Based on the models with Student-t errors, the implied quantile level that the 1\% ES is estimated to fall at is $\approx$ 0.36\%.
If that is accurate, then the CARE and ES-CAViaR-AR have a conservative ES violation rate, while the ES-CAViaR-AR-SSRR model is slightly anti-conservative (none are significantly different to 0.36\% by the UC test).

However, through closer inspection of Figure \ref{Fig_es_fore_zoom_in}, the cost efficiency gains from the ES-CAViaR-AR-SSRR model are
again observed, in a similar pattern to that from the VaR forecasting study. The CARE model achieves a lower than nominal ES VRate by generating relatively more extreme ES forecasts than the ES-CAViaR-AR-SSRR model's on 1738 days (82\%). In addition, the original ES-CAViaR-AR model is more extreme than ES-CAViaR-AR-SSRR on 1668 days (79\%).

Therefore, the ES-CAViaR-AR-SSRR can improve the forecast efficiency and lead to lower capital allocations to protect against extreme returns, compared with the CARE and ES-CAViaR-AR, while still achieving an acceptable violation rate. Again, such extra efficiency is also frequently observed for
the ES-CAViaR-X type models in the other time series.

\subsubsection{\normalsize VaR\&ES Joint Loss Function}

In this section, a joint VaR\&ES loss function (Fissler and Ziegel, 2016) is employed to compare the models VaR and ES forecasts jointly, and to
help clarify and quantify any extra efficiency gains from the ES-CAViaR-X ES forecasts compared to its competitors.

Fissler and Ziegel (2016) developed a family of loss functions that are a joint function of the associated VaR and ES series. This loss
function family is strictly consistent for the true VaR and ES series, i.e. uniquely minimized by the true VaR and ES series. The general
function family form is:
\begin{eqnarray*}
S_t(r_t, VaR_t, ES_t) &=& (I_t -\alpha)G_1(VaR_t) - I_tG_1(r_t) +  G_2(ES_t)\left(ES_t-VaR_t + \frac{I_t}{\alpha}(VaR_t-r_t)\right) \\
                      &-& H(ES_t) + a(r_t) \, ,
\end{eqnarray*}
where $I_t=1$ if $r_t<VaR_t$ and 0 otherwise for $t=1,\ldots,T$, $G_1()$ is increasing, $G_2()$ is strictly increasing and strictly convex,
$G_2 = H^{'}$ and $\lim_{x\to -\infty} G_2(x) = 0$ and $a(\cdot)$ is a real-valued integrable function.

As discussed in Taylor (2017),  making the choices: $G_1(x) =0$, $G_2(x) = -1/x$, $H(x)= -\text{log}(-x)$ and  $a= 1-\text{log} (1-\alpha)$,
which satisfy the required criteria, returns the scoring function:
\begin{eqnarray}\label{es_caviar_log_score}
S_t(r_t, VaR_t, ES_t) = -\text{log} \left( \frac{\alpha-1}{\text{ES}_t} \right) - {\frac{(r_t-Q_t)(\alpha-I(r_t\leq Q_t))}{\alpha \text{ES}_t}},
\end{eqnarray}

where the loss function is $S = \sum_{t-1}^T S_t$. Taylor (2017) referred to expression (\ref{es_caviar_log_score}) as the AL log score.
Function (\ref{es_caviar_log_score}) is exactly the negative of the AL log-likelihood in Equation (\ref{es_caviar_like_equation}), and is a
strictly consistent scoring rule that is jointly minimized by the true VaR and ES series. We use this to informally and jointly assess and compare the VaR and ES forecasts from all models.

Tables \ref{veloss} and \ref{veloss_summary} shows the loss function values $S$, calculated using Equation (\ref{es_caviar_log_score}),
which jointly assesses the accuracy of each model's VaR and ES series, during the forecast period for each market. On this measure, the
ES-CAViaR-X models employing sub-sampled RV and sub-sampled RR do best overall, having lower loss than most other models in most series and being consistently ranked better. Another observation here is that the VaR \& ES joint loss from ES-CAViaR-Exp-X specification is in general slightly
lower than that of ES-CAViaR-AR-X specification. 
The G-t model ranks lowest among individual models.

Generally the ES-CAViaR-X models are better ranked with lower loss than other models in most markets. These models consistently outperform the
all other models.

\begin{table}[!ht]
\begin{center}
\caption{\label{veloss} \small VaR and ES joint loss function values across the markets; $\alpha=0.01$.}\tabcolsep=10pt
\tiny
\begin{tabular}{lccccccccccccc} \hline
Model     &S\&P 500         &NASDAQ         &HK             &FTSE           &DAX            &SMI            &ASX200         &IBM            &GE    \\ \hline
G-t&4794.95&5067.21&5144.41&\cb{4872.06}&\cred{5285.21}&\cred{4987.11}&4531.87&5818.50&5687.71\\
EG-t&4800.92&5068.73&4985.20&4837.59&5277.27&4905.24&4503.69&\cred{5844.20}&5681.00\\
GJR-t    &4665.49&4967.73&5009.91&4793.77&\bf{5315.81}&\bf{4993.55}&4475.17&\bf{5875.65}&5668.70\\
Gt-HS    &4768.79&5031.32&5100.87&4811.84&5274.46&4884.34&4510.39&5807.27&5695.32\\
CARE          &\cred{4836.65}&\bf{5201.47}&5018.52&\bf{4890.94}&5231.81&4973.17&\bf{4793.38}&5650.35&\bf{6128.27}\\
RG-RV-GG   &4706.04&4948.34&\bf{5673.09}&4768.50&5275.24&4878.01&4432.72&5840.04&5436.72\\
RG-RV-tG      &4590.75&\fbox{4875.38}&\cred{5288.32}&4706.45&5146.03&4778.00&4386.38&5583.01&5440.95\\
ES-CAV-AR&\bf{4844.03}&\cred{5099.68}&5069.05&4859.62&5237.14&4941.23&\cred{4596.89}&5620.01&\cred{5752.51}\\
ES-CAV-AR-RV&4553.38&4979.85&5283.14&4733.96&5161.25&4765.70&4393.98&5655.13&5450.24\\
ES-CAV-AR-RR&4512.77&4904.81&5130.03&4701.78&5136.63&4681.40&4471.25&5667.03&5402.04\\
ES-CAV-AR-ScRV&4582.19&4993.35&5082.39&4698.84&5200.58&4751.53&4422.27&5635.32&5414.89\\
ES-CAV-AR-ScRR&4566.82&4966.42&\fbox{4960.45}&4719.16&5157.52&\cb{4661.10}&4489.34&\fbox{5607.45}&5525.87\\
ES-CAV-AR-SSRV&4532.31&4934.92&4986.46&\fbox{4670.50}&5133.59&4715.00&4422.38&5641.66&5498.74\\
ES-CAV-AR-SSRR&\cb{4499.02}&4892.78&5063.14&4681.19&5123.04&4713.11&4434.03&5637.23&5465.67\\
ES-CAV-Exp&4833.55&5068.22&5071.51&4875.55&5234.80&4909.00&4589.51&5622.74&5707.62\\
ES-CAV-Exp-RV&4542.91&4968.21&5285.80&4727.42&5155.61&4745.98&\fbox{4391.79}&5640.00&\cb{5401.63}\\
ES-CAV-Exp-RR&4496.48&4902.64&5123.30&4699.49&5126.99&4685.42&4463.13&5661.98&\fbox{5398.13}\\
ES-CAV-Exp-ScRV&4581.15&4978.76&5080.18&4696.85&5196.69&4742.93&4422.98&5642.86&5439.76\\
ES-CAV-Exp-ScRR&4555.45&4962.63&\cb{4965.12}&4712.26&5146.90&\fbox{4645.23}&4491.83&\cb{5619.60}&5473.15\\
ES-CAV-Exp-SSRV&4515.19&4922.43&4986.50&\cb{4671.12}&\fbox{5116.57}&4699.77&\cb{4419.92}&5628.74&5460.06\\
ES-CAV-Exp-SSRR&\fbox{4488.89}&\cb{4892.04}&5060.51&4675.51&\cb{5122.24}&4702.79&4430.34&5644.69&5479.05\\
\hline
\end{tabular}
\end{center}
\emph{Note}:\small  Box indicates the favoured model, blue shading indicates the 2nd ranked model, bold indicates the least favoured model,
red shading indicates the 2nd lowest ranked model, in each column.
\end{table}


\begin{table}[!ht]
\begin{center}
\caption{\label{veloss_summary} \small VaR and ES joint loss function values summary; $\alpha=0.01$.}\tabcolsep=10pt
\begin{tabular}{lccccccccccccc} \hline
Model      & Mean loss          & Mean rank  \\ \hline
G-t&\cred{5132.11}&\bf{18.11}\\
EG-t&5100.43&16.00\\
GJR-t    &5085.09&15.22\\
Gt-HS    &5098.29&16.22\\
CARE          &\bf{5191.62}&\cred{17.44}\\
RG-RV-GG   &5106.52&13.67\\
RG-RV-tG      &4977.25&8.00\\
ES-CAV-AR&5113.35&16.33\\
ES-CAV-AR-RV&4997.40&11.22\\
ES-CAV-AR-RR&4956.42&8.11\\
ES-CAV-AR-ScRV&4975.70&9.56\\
ES-CAV-AR-ScRR&4961.57&8.22\\
ES-CAV-AR-SSRV&4948.39&6.67\\
ES-CAV-AR-SSRR&4945.47&6.33\\
ES-CAV-Exp&5101.39&15.89\\
ES-CAV-Exp-RV&4984.37&9.11\\
ES-CAV-Exp-RR&4950.84&7.00\\
ES-CAV-Exp-ScRV&4975.80&9.56\\
ES-CAV-Exp-ScRR&4952.46&7.56\\
ES-CAV-Exp-SSRV&\fbox{4935.59}&\fbox{4.78}\\
ES-CAV-Exp-SSRR&\cb{4944.01}&\cb{6.00}\\
\hline
\end{tabular}
\end{center}
\emph{Note}:\small  Boxes indicate the favoured model, blue shading indicates the 2nd ranked model, bold indicates the least favoured model,
red shading indicates the 2nd lowest ranked model, in each column. "Mean rank" is the average rank across the 7 markets and 2 assets for the
loss function, over the 21 models: lower is better.
\end{table}

\subsubsection{\normalsize Model Confidence Set}

The model confidence set (MCS) was introduced by Hansen, Lunde and Nason (2011), as a method to statistically compare a
group of forecast models via a loss function. We apply MCS to further compare among the 21 (VaR, ES) forecasting models.
A MCS is a set of models constructed such that it will contain the
best model with a given level of confidence, which was selected as 90\% in our paper. The Matlab code for MCS testing was
downloaded from "www.kevinsheppard.com/MFE\_Toolbox". We adapted code to incorporate the VaR and ES joint loss function values
(Equation (\ref{es_caviar_log_score})) as the loss function during the MCS calculation. Each of two methods (R and SQ) to calculate the test statistics are employed in the MCS selection process. The main difference of R method and
SQ method is the way of calculating the test statistics, one uses absolute values summed during the calculation ('R'), and one uses
squares summed ('SQ'), details as in page 465 of Hansen \emph{et al.} (2011).

Table \ref{mcs_r} and \ref{mcs_sq} present the 90\% MCS using the R and SQ methods, respectively. Column "Total" counts the
total number of times that a model is included in the 90\% MCS across the 9 return series. Based on this column, boxes indicate the
favoured model, and blue shading indicates the 2nd ranked model. Bold indicates the least favoured and red shading
indicates the 2nd lowest ranked model.

Via the R method, 8 ES-CAViaR-X type models are included in the MCS for all markets and assets, and the rest 4 ES-CAViaR-X type models are included in the MCS for 8 times, together with EG-t and RG-RV-tG.
The original ES-CAViaR-AR and ES-CAViaR-Exp are included the MCS for only 4 and 5 times respectively. Via the SQ method, all 12 proposed ES-CAViaR models are included in the MCS for all 9 series.

\begin{table}[!ht]
\begin{center}
\caption{\label{mcs_r} \small 90\% model confidence set with R method across the markets and assets.}\tabcolsep=10pt
\tiny
\begin{tabular}{lccccccccccccc} \hline
Model    &S\&P 500  &NASDAQ &HK  &FTSE  &DAX  &SMI   &ASX200 &IBM &GE  &Total  \\ \hline
G-t&0&1&0&1&1&0&1&1&0&\cred{5}\\
EG-t&1&1&1&1&1&1&1&1&0&\cb{8}\\
GJR-t    &1&1&1&1&1&1&1&0&0&7\\
Gt-HS    &0&1&1&1&1&1&1&0&1&7\\
CARE          &0&1&1&1&1&0&0&1&0&\cred{5}\\
RG-RV-GG   &0&1&0&1&1&1&1&1&1&7\\
RG-RV-tG      &1&1&0&1&1&1&1&1&1&\cb{8}\\
ES-CAV-AR&0&0&1&1&1&0&0&1&0&\bf{4}\\
ES-CAV-AR-RV&1&1&0&1&1&1&1&1&1&\cb{8}\\
ES-CAV-AR-RR&1&1&1&1&1&1&1&1&1&\fbox{9}\\
ES-CAV-AR-ScRV&1&1&1&1&1&1&1&1&1&\fbox{9}\\
ES-CAV-AR-ScRR&1&1&1&1&1&1&0&1&1&\cb{8}\\
ES-CAV-AR-SSRV&1&1&1&1&1&1&1&1&1&\fbox{9}\\
ES-CAV-AR-SSRR&1&1&1&1&1&1&1&1&1&\fbox{9}\\
ES-CAV-Exp&0&1&1&1&1&0&0&1&0&\cred{5}\\
ES-CAV-Exp-RV&1&1&0&1&1&1&1&1&1&\cb{8}\\
ES-CAV-Exp-RR&1&1&1&1&1&1&1&1&1&\fbox{9}\\
ES-CAV-Exp-ScRV&1&1&1&1&1&1&1&1&1&\fbox{9}\\
ES-CAV-Exp-ScRR&1&1&1&1&1&1&0&1&1&\cb{8}\\
ES-CAV-Exp-SSRV&1&1&1&1&1&1&1&1&1&\fbox{9}\\
ES-CAV-Exp-SSRR&1&1&1&1&1&1&1&1&1&\fbox{9}\\
\hline
\end{tabular}
\end{center}
\emph{Note}:\small Boxes indicate the favoured model, blue shading indicates the 2nd ranked model, bold indicates the least favoured model,
red shading indicates the 2nd lowest ranked model, based on total number of included in the MCS across the 7 markets and 2 assets, higher is better.
\end{table}

\begin{table}[!ht]
\begin{center}
\caption{\label{mcs_sq} \small 90\% model confidence set with SQ method across the markets and assets.}\tabcolsep=10pt
\tiny
\begin{tabular}{lccccccccccccc} \hline
Model    &S\&P 500  &NASDAQ &HK  &FTSE  &DAX  &SMI   &ASX200 &IBM &GE  &Total  \\ \hline
G-t&0&1&1&1&1&0&1&1&1&7\\
EG-t&0&1&1&1&1&1&1&1&0&7\\
GJR-t    &0&1&1&1&1&1&1&1&0&7\\
Gt-HS    &0&1&1&1&1&1&1&1&1&\cb{8} \\
CARE          &0&1&1&1&1&0&0&1&0& \bf{5}\\
RG-RV-GG   &0&1&0&1&1&1&1&1&1&7\\
RG-RV-tG      &1&1&1&1&1&1&1&1&1&\fbox{9}\\
ES-CAV-AR&0&1&1&1&1&0&1&1&0&\cred{6}\\
ES-CAV-AR-RV&1&1&1&1&1&1&1&1&1&\fbox{9}\\
ES-CAV-AR-RR&1&1&1&1&1&1&1&1&1&\fbox{9}\\
ES-CAV-AR-ScRV&1&1&1&1&1&1&1&1&1&\fbox{9}\\
ES-CAV-AR-ScRR&1&1&1&1&1&1&1&1&1&\fbox{9}\\
ES-CAV-AR-SSRV&1&1&1&1&1&1&1&1&1&\fbox{9}\\
ES-CAV-AR-SSRR&1&1&1&1&1&1&1&1&1&\fbox{9}\\
ES-CAV-Exp&0&1&1&1&1&0&1&1&1&7\\
ES-CAV-Exp-RV&1&1&1&1&1&1&1&1&1&\fbox{9}\\
ES-CAV-Exp-RR&1&1&1&1&1&1&1&1&1&\fbox{9}\\
ES-CAV-Exp-ScRV&1&1&1&1&1&1&1&1&1&\fbox{9}\\
ES-CAV-Exp-ScRR&1&1&1&1&1&1&1&1&1&\fbox{9}\\
ES-CAV-Exp-SSRV&1&1&1&1&1&1&1&1&1&\fbox{9}\\
ES-CAV-Exp-SSRR&1&1&1&1&1&1&1&1&1&\fbox{9}\\
\hline
\end{tabular}
\end{center}
\emph{Note}:\small Boxes indicate the favoured model, blue shading indicates the 2nd ranked model, bold indicates the least favoured model,
red shading indicates the 2nd lowest ranked model, based on total number of included in the MCS across the 7 markets and 2 assets, higher is better.
\end{table}

Overall, across several measures and test for forecasts accuracy and model comparison, when forecasting 1\% VaR and ES in 9 financial
return series, the ES-CAViaR-X type models generally performed in a highly
favourable manner when compared to a range of competing models. When considering VRates, rejections by standard tests, quantile loss and VaR\&ES joint loss, the ES-CAViaR-AR-X and ES-CAViaR-Exp-X models employing sub-sampled RV and sub-sampled RR are the most favourable models overall.

{\centering
\section{\normalsize CONCLUSION}\label{conclusion_section}
\par
}
\noindent
In this paper, the original ES-CAViaR models are extended through incorporating
intra-day and high frequency volatility measures, to estimate and forecast financial tail risk. Improvements in the out-of-sample forecasting of tail risk measures are observed, compared to Re-GARCH models employing realized volatility, and
traditional GARCH and CARE models, as well as the original ES-CAViaR models. Specifically, ES-CAViaR-AR-X and ES-CAViaR-Exp-X models employing sub-sampled RV and  sub-sampled RR generate the best VaR and Es forecasting results in the empirical study of 9 financial return series. 
With respect to the back testing of VaR forecasts, the ES-CAViaR-X type models are
also generally less likely to be rejected than their counterparts.
Further, the model confidence set results also apparently favour the proposed ES-CAViaR-X framework.
In addition to being more accurate, the ES-CAViaR-X models generated less extreme tail risk forecasts, regularly allowing smaller amounts of capital allocation without being anti-conservative or significantly inaccurate.

To conclude, the ES-CAViaR-X type models, especially the ones use sub-sampled RV and sub-sampled RR, should be considered for
financial applications when forecasting tail risk, and should allow financial institutions to more accurately allocate capital under the
Basel Capital Accord, to protect their investments from extreme market movements. This work could be extended by developing asymmetric and non-linear CAViaR specifications; by improving ES component of the model; by using
alternative frequencies of observation for the realized measures and by extending the framework to allow multiple realized measures to
appear simultaneously in the model.

\clearpage
\section*{References}
\addcontentsline{toc}{section}{References}
\begin{description}
\item Andersen, T. G. and Bollerslev, T. (1998). Answering the skeptics: Yes, standard volatility models do provide accurate
forecasts. \emph{International economic review}, 885-905.

\item Andersen, T. G., Bollerslev, T., Diebold, F. X. and Labys, P. (2003). Modeling and forecasting realized volatility.
    \emph{Econometrica}, 71(2), 579-625.

\item Artzner, P., Delbaen, F., Eber, J.M., and Heath, D. (1997). Thinking coherently.  \emph{Risk}, 10, 68-71.

\item Artzener, P., Delbaen, F., Eber, J.M., and Heath, D. (1999). Coherent measures of risk.  \emph{Mathematical Finance}, 9(3), 203-228.



\item Chen, W., Peters, G., Gerlach, R. and Sisson, S. (2017). Dynamic Quantile Function Models. arXiv:1707.02587.

\item Christensen, K. and Podolskij, M. (2007). Realized range-based estimation of integrated variance. \emph{Journal of Econometrics},
141(2), 323-349.

\item Christoffersen, P. (1998). Evaluating interval forecasts. \emph{International Economic Review}, 39, 841-862.


\item Damien, P., Dellaportos, P. Polson, N. and Stephens, D. (2013). {\em Bayesian Theory and Applications}, pg 99, Oxford University Press


\item Engle, R. F. and Manganelli, S. (2004). CAViaR: Conditional Autoregressive Value at Risk
by Regression Quantiles. \emph{Journal of Business and Economic Statistics}, 22(4), 367-381.

\item Feller, W. (1951). The Asymptotic Distribution of the Range of Sums of Random Variables. \emph{Annals of Mathematical Statistics},
    22, 427-32.

\item Fissler, T. and Ziegel, J. F. (2016). Higher order elicibility and Osband's principle. \emph{Annals of Statistics}, 44(4), 1680-1707.

\item Gaglianone, W. P., Lima, L. R., Linton, O. and Smith, D. R. (2011). Evaluating Value-
at-Risk models via quantile regression. \emph{Journal of Business \& Economic Statistics},
29(1), 150-160.

\item Garman, M. B. and Klass, M. J. (1980). On the Estimation of Security Price Volatilities from historical data.
\emph{The Journal of Business}, 67-78.

\item Gerlach, R. and Chen, C.W.S. (2016). Bayesian Expected Shortfall Forecasting Incorporating the Intraday Range,
\emph{Journal of Financial Econometrics}, 14(1), 128–158.

\item Gerlach, R., Chen, C.W.S. and Chan, N.Y. (2011). Bayesian time-varying quantile forecasting for value-at-risk in financial markets. \emph{Journal of Business \& Economic Statistics}, 29(4), 481-492.



\item Gerlach, R. and Wang, C. (2016).  Forecasting risk via realized GARCH, incorporating the realized range.
\emph{Quantitative Finance}, 16(4), 501-511.

\item Gneiting, T (2011). Making and evaluating point forecasts. \emph{Journal of the American Statistical Association}, 106(494), 746-762.



\item Hansen, P. R., Huang, Z. and Shek, H. H. (2011). Realized GARCH: a joint model for returns and realized measures of volatility.
\emph{Journal of Applied Econometrics}, 27(6), 877-906.

\item Hansen, P.R., Lunde, A. and Nason, J.M. (2011). The model confidence set. Econometrica, 79(2), 453-497.

\item Harvey, A.C. and T. Chakravarty. (2009). Beta-t-(E)GARCH. Cambridge Working Papers in Economics 0840, Faculty of Economics, University of Cambridge, Cambridge.


\item Koenker, R. and Machado, J.A. (1999). Goodness of fit and related inference processes for quantile regression. \emph{Journal of the American Statistical Association}, 94(448), 1296-1310.

\item Kupiec, P. H. (1995). Techniques for Verifying the Accuracy of Risk Measurement Models. \emph{The Journal of Derivatives}, 3, 73-84.

\item Martens, M. and van Dijk, D. (2007). Measuring volatility with the realized range. \emph{Journal of Econometrics}, 138(1), 181-207.


\item Moln{\'a}r, P. (2012). Properties of range-based volatility estimators. \emph{International Review of Financial Analysis}, 23,
    20-29.

\item Metropolis, N., Rosenbluth, A. W., Rosenbluth, M. N., Teller, A. H., and Teller, E. (1953). Equation of State Calculations by Fast
Computing Machines. \emph{The journal of chemical physics}, 21(6),1087-1092.


\item Parkinson, M. (1980). The extreme value method for estimating the variance of the rate of return. \emph{Journal of Business},
53(1), 61.

\item Rogers, L. C. G. and Satchell, S. E. (1991). Estimating variance from high, low and closing prices. \emph{Annals of Applied
    Probability}, 1, 504-512.

\item Roberts, G. O., Gelman, A. and Gilks, W. R. (1997). Weak convergence and optimal scaling of random walk Metropolis algorithms.
    \emph{The annals of applied probability}, 7(1), 110-120.

\item Schwert, G.W. (1989). Why Does Stock Market Volatility Change Over Time? \emph{Journal of Finance}, 44, 1115-1153.



\item Taylor, J. (2017). Forecasting value at risk and expected shortfall using a semiparametric approach based on the asymmetric Laplace distribution. \emph{Journal of Business \& Economic Statistics}, 1-13.

\item Taylor, S.J. (1986). Modeling Financial Time Series. Chichester, UK: John Wiley and Sons.


\item Yang, D. and Zhang, Q. (2000). Drift-independent volatility estimation based on high, low, open, and close prices. \emph{Journal of
    Business}, 73, 477-491.

\item Zhang, L., Mykland, P. A., and A\"{i}t-Sahalia, Y. (2005). A tale of two time scales.  \emph{Journal of the American Statistical
    Association}, 100(472), 1394-1411.

\end{description}

\clearpage
\appendixtitleon
\appendixtitletocon
\begin{appendices}

{\centering
\section{\normalsize ES-CAViaR SIMULATION STUDY}\label{simulation_return_section}
\par
}
To study performance of the performance of the Bayesian method and MLE for the ES-CAViaR type models (models (\ref{es_caviar_ar_model}) and (\ref{es_caviar_exp_model})), another two simulation studies were conducted, under a similar manner to the ones in Section \ref{simulation_section}. The difference is that an absolute value GARCH model without realized measures, as in model (\ref{r_garch_simu}), is used as the data generation process.

1000 replicated datasets of size $n=1900$ are simulated from the following specific absolute value GARCH model. Again, the equivalent ES-CAViaR model was fit to each data set, once using MCMC and once using ML.

\begin{eqnarray} \label{r_garch_simu}
&&r_t= \sqrt{h_t} \epsilon_t^{*} \\ \nonumber
&&\sqrt{h_t}= 0.02 + 0.10 |r_{t-1}|+ 0.85 \sqrt{h_{t-1}}  \\ \nonumber
&& \epsilon_t^{*} \stackrel{\rm i.i.d.} {\sim} N(0,1)\\ \nonumber
\end{eqnarray}

A mapping between from the absolute value GARCH (Abs-GARCH) to the ES-CAViaR, the "true" $\gamma_0, \gamma_1, \gamma_2$ under ES-CAViaR-AR and "true" $\gamma_0$ uner ES-CAViaR-Exp are all calculated under the same way as in Section  \ref{simulation_section}.

Estimation results of ES-CAViaR-AR are summarized in Table \ref{simu_table}. First, accurate parameter estimates and VaR \& ES forecasting results are produced by both adaptive MCMC and ML, which proves the validity of both methods for the ES-CAViaR-AR models. Further, bias the results favour the MCMC estimator compared to the MLE for 5 out of 6 parameter and for VaR \& ES forecasts. Further, the precision is higher for the MCMC method
for 4 of 6 parameters and for both VaR and ES forecasts.

With respect to the ES-CAViaR-Exp model estimation, we have a slight different story. Both MCMC and ML still generate relatively accurate parameter estimates and VaR \& ES forecasts in this case. However,
the MCMC and ML methods produce quite close performance. MCMC produces better results regarding bias, for 3 out of 4 parameters and for VaR forecast. The precision results favoured ML for 3 out of 4 parameters and for ES forecast. As discussed in Section \ref{simulation_section}, ES-CAViaR-Exp has a relatively simpler framework, thus the estimation is less challenge to both MCMC and ML, while for more complicated framework, e.g. ES-CAViaR-AR, MCMC clearly demonetized its advantageous.

\begin{table}[!ht]
\begin{center}
\caption{\label{simu_table} \small Summary statistics for the two estimators of the ES-CAViaR-AR model, with data simulated from model (\ref{r_garch_simu}).}\tabcolsep=10pt

\begin{tabular}{lcccccccc} \hline
$n=1900$               &             & \multicolumn{2}{c}{MCMC}      &  \multicolumn{2}{c}{ML}   \\
Parameter              &True         &Mean           &  RMSE         &Mean           & RMSE    \\ \hline
$\beta_0$         &       -0.0465 &	\fbox{-0.0667} &	 0.0987 &	-0.0792 &	 \fbox{0.0771}     \\
$\beta_1$         &      -0.2326 	& \fbox{-0.2593} 	& \fbox{0.1013} &	-0.2741 	& 0.1076   \\
$\beta_2$          &    0.8500 &	 \fbox{0.8102} &	 0.1628 &	 0.7861 &	 \fbox{0.1352}    \\
$\gamma_0$          &     0.0550 	& \fbox{0.0442} 	& \fbox{0.0341} &	 0.0409 &	 0.0461  \\
$\gamma_1$           &    0.1822  &	 0.2404 	& \fbox{0.2271} &	 \fbox{0.1428} &	 0.2671  \\
$\gamma_2$            &    0.2554 	& \fbox{0.2758} 	& \fbox{0.2567} 	& 0.3849 	& 0.4504   \\
$\text{VaR}_{n+1}$     &        -0.6597& 	\fbox{-0.6599} &	 \fbox{0.0440} &	-0.6593 	& 0.0457    \\
$\text{ES}_{n+1}$     &      -0.7559 	&\fbox{-0.7470} &	 \fbox{0.0537} &	-0.7381& 	 0.0592      \\
\hline
\end{tabular}
\end{center}
\emph{Note}:\small  A box indicates the favored estimators, based on mean and RMSE.
\end{table}

\begin{table}[!ht]
\begin{center}
\caption{\label{simu_table_1} \small Summary statistics for the two estimators of the ES-CAViaR-Exp model, with data simulated from model (\ref{r_garch_simu}).}\tabcolsep=10pt
\begin{tabular}{lcccccccc} \hline
$n=1900$               &             & \multicolumn{2}{c}{MCMC}      &  \multicolumn{2}{c}{ML}   \\
Parameter              &True         &Mean           &  RMSE         &Mean           & RMSE    \\ \hline
$\beta_0$         &   -0.0465 	&\fbox{-0.0747} &	 0.1218 &	-0.0805 	& \fbox{0.0750}       \\
$\beta_1$         &     -0.2326 &	\fbox{-0.2580} 	& \fbox{0.0989}& 	-0.2768 &	 0.1067  \\
$\beta_2$          &      0.8500 &	 \fbox{0.7994} 	& 0.1878 &	 0.7834 	& \fbox{0.1331}    \\
$\gamma_0$          &    -1.9264 &	-2.1236 	& 0.3668 &	\fbox{-2.0210} 	& \fbox{0.2541}    \\
$\text{VaR}_{n+1}$     &    -0.6583 &	\fbox{-0.6588} &	 \fbox{0.0432} 	&-0.6574 	& 0.0433       \\
$\text{ES}_{n+1}$ s     &      -0.7542& 	-0.7434 &	 0.0514 &	\fbox{-0.7466} &	 \fbox{0.0507}  \\
\hline
\end{tabular}
\end{center}
\emph{Note}:\small  A box indicates the favored estimators, based on mean and RMSE.
\end{table}

\end{appendices}

\end{document}